# Strong plasmon reflection at nanometer-size gaps in monolayer graphene on SiC

Jianing Chen[1], Maxim L. Nesterov[2], Alexey Yu. Nikitin[1,2], Sukosin Thongrattanasiri[4,5], Pablo Alonso-González[1], Tetiana M. Slipchenko[2], Florian Speck[6], Markus Ostler[7], Thomas Seyller[7], Iris Crassee[8], Frank H. L Koppens[9], Luis Martin-Moreno[2], F. Javier García de Abajo [4,9], Alexey B. Kuzmenko[8], Rainer Hillenbrand[1,3],★

[1] CIC nanoGUNE Consolider, 20018 Donostia-San Sebastián, Spain
[2] Instituto de Ciencia de Materiales de Aragón and Departamento de Física de la Materia Condensada, CSIC-Universidad de Zaragoza, E-50009, Zaragoza, Spain
[3] IKERBASQUE Basque Foundation for Science, 48011 Bilbao, Spain
[4] ICREA-Institució Catalana de Recerca i Estudis Avançats, Barcelona, Spain
[5] Department of Physics, Kasetsart University, Bangkok 10900, Thailand
[6] Department Physik, Universität Erlangen-Nürnberg, 91058 Erlangen, Germany
[7] Institut für Physik - Technische Physik, Technische Universität Chemnitz, 09126 Chemnitz, Germany
[8] Département de Physique de la Matie re Condensée, Université de Genève, 1211 Genève, Switzerland
[9] ICFO-Institut de Ciéncies Fotoniques, Mediterranean Technology Park, 08860 Castelldefels, Barcelona, Spain
★ Email: r.hillenband@nanogune.eu

**We employ tip-enhanced infrared near-field microscopy to study the plasmonic properties of epitaxial quasi-free-standing monolayer graphene on silicon carbide. The near-field images reveal propagating graphene plasmons, as well as a strong plasmon reflection at gaps in the graphene layer, which appear at the steps between the SiC terraces. When the step height is around 1.5 nm, which is two orders of magnitude smaller than the plasmon wavelength, the reflection signal reaches 20% of its value at graphene edges, and it approaches 50% for step heights as small as 5 nm. This intriguing observation is corroborated by numerical simulations, and explained by the accumulation of a line charge at the graphene termination. The associated electromagnetic fields at the graphene termination decay within a few nanometers, thus preventing efficient plasmon transmission across nanoscale gaps. Our work suggests that plasmon propagation in graphene-based circuits can be tailored using extremely compact nanostructures, such as ultra-narrow gaps. It also demonstrates that tip-enhanced near-field microscopy is a powerful contactless tool to examine nanoscale defects in graphene.**

Graphene plasmons are electromagnetic waves propagating along graphene layers[1-10]. They exhibit a remarkable electrostatic tunability and the ability to strongly concentrate electromagnetic energy, potentially leading to new subwavelength-scale plasmonic and optoelectronic applications. One of the outstanding challenges towards the realization of graphene-based plasmonic circuits is the efficient on-chip manipulation of the plasmonic energy flow in CVD-grown graphene[11,12] or in epitaxial graphene on SiC[13-16]. It has been recently reported that graphene plasmons are strongly reflected not only at graphene edges[8,9] but also at defects of extremely subwavelength scale dimensions[17] and at grain boundaries[18]. While uncontrolled plasmon reflections at naturally grown defects may represent major obstacles for the development of graphene plasmonic devices, the controlled structuring of graphene on the 1 nm scale may open in the near future promising avenues for steering plasmons with ultracompact reflecting elements.

Regardless of a particular future technological realization, it is interesting from both fundamental and applied perspectives to study how efficient plasmons are reflected when the critical dimensions of reflecting elements are much smaller than the plasmon wavelength. Here we address this question by measuring plasmon reflection from nanometer-size gaps, which are expected to form at terrace steps in so-called quasi-free-standing monolayer graphene (QFMLG)[15,19] on the Si-face of SiC. QFMLG is produced by forming a so-called buffer layer with $(6\sqrt{3}\times6\sqrt{3})R30°$ periodicity by annealing of the SiC surface, followed by decoupling of that layer through intercalation of hydrogen. This preparation results in significant p-type doping ($\sim 6\times10^{12}$ cm$^{-2}$)[19-21], which is essential for the existence of well-defined propagating plasmon modes. More information about the sample preparation is provided in the Methods section. The substrate terraces are atomically flat resulting in high-quality homogeneous graphene areas with a lateral extension of up to a few micrometers. The substrate terraces are separated by nanometer-size steps, which suggests that QFMLG is an interesting system to study graphene plasmon reflection at ultrasmall discontinuities (gaps in the graphene layer) without the need of patterning. Indeed, an intense terahertz absorption peak was recently observed in QFMLG[22], indicating a strong influence of terrace steps on its far-field optical properties. In this paper we use scattering-type scanning near-field optical microscopy (s-SNOM)[23,24], which was recently employed for interferometric plasmon imaging at graphene edges,[8,9] to directly observe plasmon reflection at the terrace steps. With this method we systematically study the plasmon reflection as a function of the step height and quantitatively analyze the data by comparing them with numerical simulations.

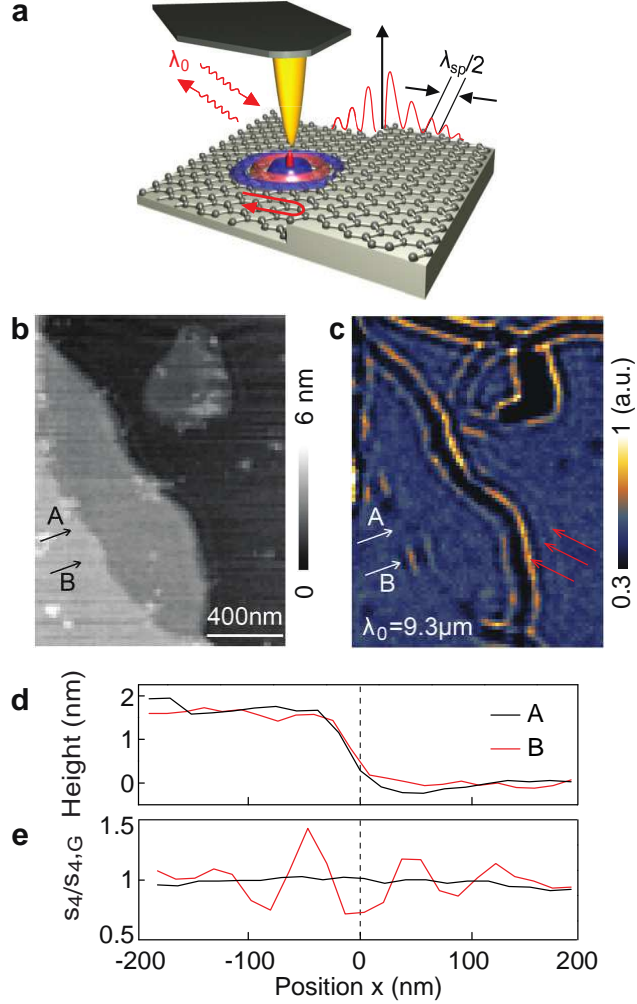

**Figure 1: Near-field imaging of graphene plasmons on SiC terraces.** a) Schematics of the s-SNOM experiment. b) AFM topography image of quasi-free standing monolayer epitaxial graphene on a 6H-SiC substrate. Red arrows indicate fringes along a substrate step. c) Optical near-field amplitude (4$^{th}$ harmonics) at $\lambda_0$ = 9.3 μm recorded in the same area. d) AFM height profile across a step at positions marked as A and B in (b) and (c). e) Normalized near-field amplitude profiles $s_4/s_{4,G}$ measured along the same line scans as in (d). $s_{4,G}$ is the near-field amplitude measured on graphene far away from the step.

The principle of plasmon imaging with s-SNOM, which is based on atomic force microscopy (AFM), is sketched in Fig. 1a. A metal-coated AFM tip is illuminated with an infrared laser of the wavelength $\lambda_0$ and the backscattered light is recorded as a function of the tip position. In order to suppress background scattering from the tip shaft and the sample, the tip is vibrated vertically and the detected signal is demodulated at a higher harmonic $n$ of the tip vibration frequency, in this work at $n = 4$. In combination with pseudoheterodyne interferometric detection scheme[24] we obtain background-free amplitude and phase signals, $s_4$ and $\varphi_4$, of which we will only use the amplitude for the sake of brevity. Importantly, the tip converts the incident light into a strongly confined near field at the tip apex, which provides the necessary momentum to launch radially emanating plasmons in a graphene layer. When these plasmons are back-reflected at edges or defects, characteristic interference patterns are observed in the near-field images.[8,9,18]

A typical topography image of one of our samples is presented in Fig. 1b (measurements are obtained at room temperature), showing terraces separated by steps of up to a few nanometers height. In the simultaneously recorded infrared near-field amplitude image (Fig. 1c) we observe several fringes (intensity minima and maxima as shown by red arrows) parallel to the steps. This bears strong resemblance to the recently discovered fringe formation near graphene edges produced by interference between the tip-emitted and back-reflected plasmons[8,9]. The observation of fringes in Fig. 1c thus indicates that plasmons are launched by the tip, which are strongly back-reflected at the terrace steps. Notably, the main fringe is brightest, as the plasmons decay rapidly in propagation direction (i.e. perpendicular to the steps). The plasmon reflection on both sides of the step is therefore hallmarked by a pair of bright parallel lines (with the step in between) accompanied by a series of weaker ones. We note that the intensity and number of fringes on the two sides can be different, as for example observed in Fig. 1c. The fringes can be more intense on either side of the step. At position B, the more intense fringe is observed at the step edge, while at the position marked by the red arrows the more intense fringe is observed at the step corner. We conclude that the asymmetry of the plasmon fringes is not related to the step geometry. It might be caused by a variation of the electronic properties (for example the carrier mobility) of the graphene monolayer among the different terraces.

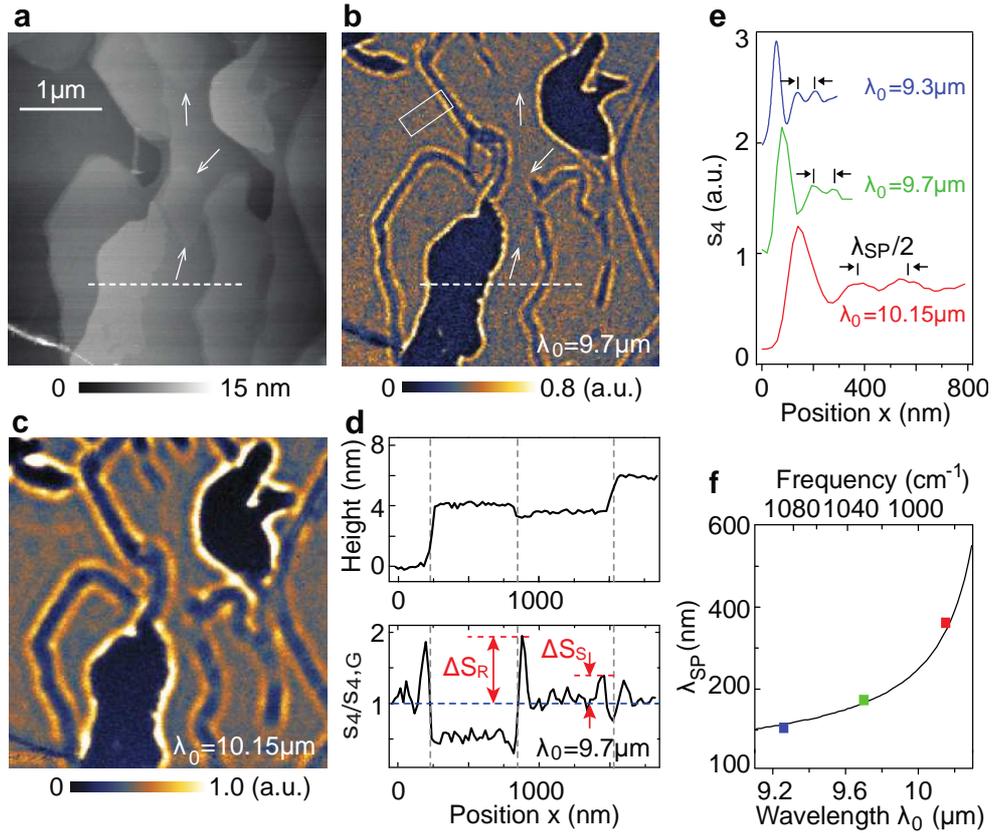

**Figure 2: Graphene plasmon dispersion on SiC substrate.** a) AFM topography image of a 4x4 µm area. Arrows mark selected low-height terrace steps (0.4 nm). b, c) Near-field amplitude images of the area shown in (a), recorded at 9.7 µm and 10.15 µm wavelength, respectively. The white rectangle in (b) indicates the position from where the near-field profiles in (e) were extracted. d) Profiles of topographic height (upper graph) and near-field amplitude (lower graph) extracted along the white dashed line in (a) and b. The vertical black dashed lines are guides to the eyes. The lower graph illustrates the definition of the values $\Delta S_R$ and $\Delta S_S$, which are used to calculate the fringe visibility $V = \Delta S_S/\Delta S_R$. e) Near-field amplitude profiles along the long side and averaged over the short side of the white rectangle in (b). f) Plasmon wavelength as a function of the incident wavelength. Squares represent experimental values extracted from (e). The solid curve shows a calculation assuming a Fermi energy of $|E_F| = 0.34$ eV.

In Fig. 2, we analyze a larger area of another sample, which contains several terraces clearly visible in the AFM topography image (Fig. 2a). Figures 2b,c show the near-field amplitude images taken at two different wavelengths (9.7 and 10.15 µm). Again, we observe fringes (most clearly the main one) parallel to the substrate steps. Notably, some terraces look much darker than the rest of the sample and do not show any plasmon interference fringes. From Raman images (not shown) we can identify these areas as graphene-free SiC. With increasing wavelength, the fringes broaden and their spacing increases. This is more obvious in Fig. 2e, where we show near-field amplitude profiles perpendicular to a step in the area marked by the white rectangle in Fig. 2b for three different wavelengths (9.3, 9.7 and 10.15 µm). Because the fringe spacing L corresponds to approximately the half of the plasmon wavelength $\lambda_{SP}$[8,9], we can plot $\lambda_{SP}$ as a function of $\lambda_0$ (symbols in Fig. 2f). The solid curve corresponds to the theoretical dispersion relation[1,4], which in the quasi-static limit ($\lambda_{SP} \ll \lambda_0$) reduces to $\lambda_{SP} = (2\lambda_0^2 e^2 |E_F|)/((11\varepsilon_{SiC}(\lambda_0))\pi c^2 \hbar^2)$, where $E_F$ is the Fermi energy with respect to the Dirac point. For the simulations we used $|E_F|$ = 0.34 eV, which is close to the experimental value[22]. $\varepsilon_{SiC}(\lambda_0)$ is the dielectric function of SiC, which is strongly wavelength dependent due to the optical phonon in SiC. Note that for the actual doping level, the interband absorption in graphene starts at much higher photon energies (at $2|E_F| \approx 0.7$ eV for vertical optical transition, but it is roughly around $|E_F|$ where the plasmon band enters the interband transitions region), and therefore we can legitimately neglect it in the above formula, which only accounts for the intraband Drude contribution. The experimental points follow rather closely the theoretical plasmon dispersion, corroborating that the fringes are a consequence of plasmon reflection at the steps.

It is noteworthy that the amplitude of the main fringe is not the same for all values of the step height, as we conclude upon inspection of both the height and normalized near-field amplitude profiles, S = $s_4$/ $s_{4,G}$, taken along the white dashed lines in Figs 2a,b, respectively (Fig. 2d). $s_{4,G}$ is the near-field amplitude on graphene measured far away from the step. We also find that the fringe amplitude at graphene edges (i.e. at the boundaries between graphene-covered and graphene-free areas, $\Delta S_R$), is always higher than the fringe amplitude at steps between graphene-covered terraces ($\Delta S_S$). Because the fringe amplitude is nearly the same for all edges, we use $\Delta S_R$ as a reference and plot the value V = $\Delta S_S/\Delta S_R$, hereafter referred to as fringe visibility, as a function of the step height $h_s$. We show below that V is indicative of the plasmon reflectivity of the steps. The result of the analysis of a large number of different steps is shown in Fig. 3c (symbols). Two observations can be immediately made. First, all steps below 1 nm show a negligible fringe visibility, such as those marked by white arrows in Fig. 2a. Second, for all steps higher than $h_s > 1.5$ nm we observe fringes, and their visibility increases from about 0.2 at $h_s$ = 1.5 nm to about 0.5 at $h_s$ = 5 nm.

Interestingly, for $h_s$ ~ 1 to 1.5 nm we find both zero and finite fringe visibility. The observation of two distinct types of fringe visibilities, either negligibly small or larger than 0.2 (indicated by black and red symbols in Fig. 3c, respectively), allows us to speculate that graphene continuously covers steps with a height below 1 nm, while for heights $h_s$ > 1.5 nm it is (electrically) disconnected. Because the fringe visibility V increases with increasing step height, we assume that a gap is formed at the step with a gap width corresponding to the step height. This assumption is supported by the recent observation that no buffer layer (and thus no graphene) is formed during the graphitization of nonpolar surfaces[25], such as the surfaces of the terrace steps.

Interestingly, we observe that some of the fringes are interrupted when the step height is about $h_s$ ~ 1.5 nm. This can be clearly seen for example at position A in Fig. 1c. Evidently, plasmons are not reflected at this step location. The step height at position A, however, is the same as at position B (Fig. 1d), where fringes indicate a strong plasmon reflection (Fig. 1e). The absence of plasmon reflection at A suggests that the graphene is continuous at this step location, thus allowing for nearly undisturbed propagation of the plasmons launched by the tip. Subsequently and in line with our observation at steps with $h_s$ > 1.5 nm, we attribute the appearance of strong fringes, for example at position B, to a discontinuity of the graphene directly at the step.

In order to test our hypotheses and to achieve a more quantitative understanding of the plasmon reflection at the terrace steps, we performed a finite elements-based simulation of the near-field contrast observed in s-SNOM. To that end, the AFM tip is approximated by a conducting ellipsoid illuminated by a plane wave. We calculated the vertical component of the electric field just below the tip, which largely determines the tip scattered field and therefore can be compared with the experimental s-SNOM amplitude (for more details see Supporting Information). The normalized field amplitude, $E_{z,norm}$, can be plotted as a function of the lateral position of the ellipsoid. The results of such calculations are shown in Figs. 3a, b. The optical conductivity of graphene, which depends on doping and the relaxation time τ, was calculated within the local (i.e., zero parallel wave vector) random-phase approximation[26-28], where we assumed $|E_F|$ = 0.34 eV and τ = 0.05 ps, corresponding to a mobility of 1430 cm$^2$/(V•s), close to the experimental values[19]. In Fig. 3a, we show the calculated near-field profile (red solid line) when the tip scans across a graphene edge (also discussed in Refs. 8 and 9), demonstrating an excellent agreement between our simulation and the experimental near-field profiles (symbols). We now apply the simulation model to compute near-field profiles perpendicular to a 1.5 nm high terrace step. We first assume that graphene does not cover the step, that is, there is a 1.5 nm wide vertical gap between two semi-infinite graphene sheets on the two terraces, as illustrated by the upper sketch in Fig. 3c. The calculated near-field profile (red solid line in Fig. 3b) exhibits strong near-field variations on both sides of the step, which resemble the experimental data (for example the red curve in Fig. 1e). In the second calculation, we assumed that graphene is continuous over the step, and for simplicity, we neglect any possible inhomogeneities near the step. The corresponding near-field profile (Fig. 3b, black curve) only exhibits mild signal variations, in agreement with the experimental data at some step positions, such as the black curve in Fig. 1e. The calculations thus confirm that the electrical connectivity of graphene at the step is the key factor determining the plasmon reflection, rather than the geometrical profile of the substrate step.

For comparison with the experimental data, we show in Fig. 3c the calculated fringe visibility $\Delta S_S/\Delta S_R$ as function of the step height $h_s$. We consider two cases. First, the graphene covers the step (as illustrated in the bottom sketch in Fig. 3c). Second, the vertical section of the step is not covered by graphene (as illustrated in the top sketch of Fig. 3c), that is, the step represents a gap in the graphene with a width corresponding to the step height. We find that for all step heights the fringe visibility for discontinuous graphene (red curve in Fig. 3c) is much higher than for continuous one (black curve in Fig. 3c). The experimentally observed increase of V with the step height is qualitatively well reproduced in the simulation, and even quantitative agreement between both is found above $h_s \approx 3$ nm. At lower step heights, the calculations yield a higher fringe visibility than what is observed in the experiment. The deviation may be due to a possible variation of the carrier density or mobility near the edge, which is not taken into account in the calculations. A future closer study of the electronic properties around the steps can probably improve the agreement between experiment and theory. However, such a study would go beyond the scope of this work and we abstain from making more specific statements in this regard.

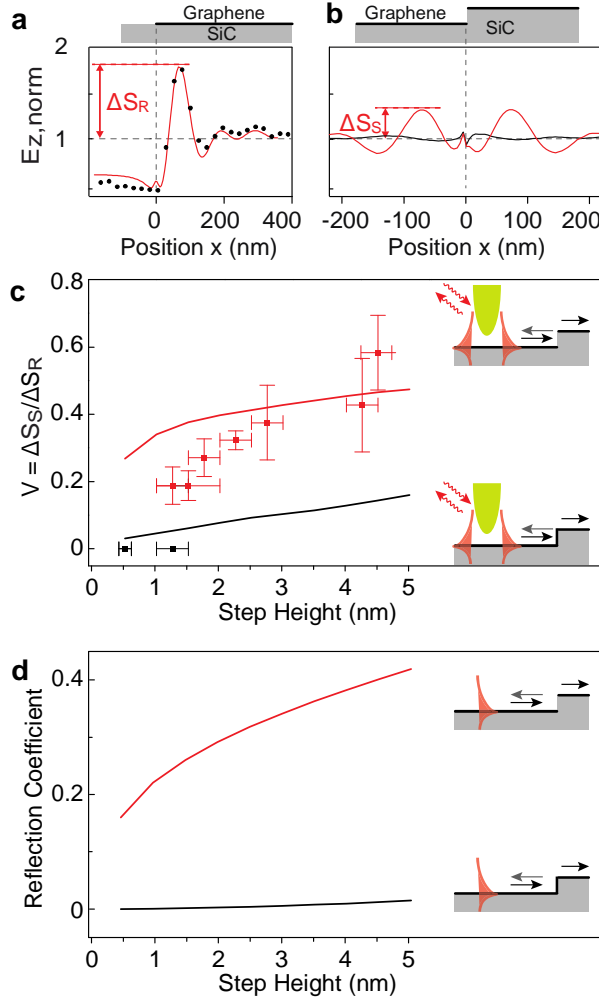

**Figure 3: Graphene plasmon reflection at SiC steps.** a) Experimental (dots) and calculated (red solid line) near-field amplitude profiles across a graphene edge, $s_4/s_{4,G}$ and, $E_{z,norm}$, respectively. A Fermi energy of $|E_F| = 0.34$ eV and a relaxation time of 0.05 ps were assumed in the calculations. b) Calculated near-field amplitude profiles, $E_{z,norm}$, for $\lambda_0 = 9.7$ μm across a 1.5 nm high step using the same parameters as in (a). The red curve shows the result obtained when graphene is disconnected at the step, whereas the black curve refers to graphene continuously covering the step. In both (a) and (b), the horizontal and vertical dashed lines mark the averaged signal on graphene and the positions of graphene edges or steps, respectively. c) Experimental and theoretical fringe visibility $V = \Delta S_S/\Delta S_R$ as a function of the step height. Each data point represents a set of measurements on various steps of similar height. Horizontal and vertical error bars correspond to the variation in measured step height and the fringe visibility, respectively. The red curve depicts the calculated V, assuming that graphene is disconnected at the step. The black curve is obtained when graphene covers the step with uniform conductivity. d) Plasmon reflection coefficient as a function of the step height for the two scenarios considered in (c), calculated analogous to ref. 17.

It is important to connect the fringe visibility observed in s-SNOM experiments and the plasmon reflection coefficient,[17] which cannot be directly measured but is essential for theoretical analysis of plasmon propagation in structured graphene. To this end, we performed another series of calculations, where we consider a plasmon in the form of a plane wave propagating along the graphene sheet towards a step (instead of a radial wave created by the tip as considered above) as it is shown in the inset of Fig. 3d. We calculate the reflection coefficient, defined as the ratio between the intensity of reflected and incident plasmons, as a function of the step height (Fig. 3d) in the same situations as above (connected and disconnected graphene at substrate steps). Similar to the fringe visibility (Fig. 3c), the plasmon reflectivity exhibits a dramatically different behavior in the two scenarios. We observe certain similarity, although not a precise match, between the step height dependence of the two quantities. Therefore, we conclude that the fringe visibility is a reliable indicator of the plasmon reflection coefficient.

The most important and striking result of our work is the experimental and theoretical demonstration that an ultranarrow graphene gap of only a few nanometers, which is almost two orders of magnitude smaller than the plasmon wavelength, can act as an efficient plasmon reflector and a reflectivity of about 50% can be achieved with gaps as small as 5 nm. To deeper rationalize this at first glance nonintuitive finding, we present in Fig. 4a and 4b the calculated spatial distribution of both x- (horizontal) and z-component of the electric fields of a plasmon wave reflecting from a graphene edge. Outside the graphene region (or inside a gap in the graphene layer), the antisymmetric field distribution of the vertical component, $E_z$, immediately cancels along x axis (Fig. 4c, black dashed curve) due to the absence of conductivity or charges. The horizontal component, $E_x$, does not cancel, however, it strongly decays within a few nanometer distance to the graphene edge along both x- and z- axes (Fig. 4c, green curves), which can be explained by the accumulation of a line charge along the extremely sharp graphene edge induced by the graphene plasmon. Beacuse of the rapid field decay in propagation direction of the plasmon, the induction of a mirror line charge in an adjacent graphene edge is strongly diminished already for gap sizes of a few nanometers. In other words, the capacitive coupling across nanometer-size graphene gaps is weak, and thus the transmission of graphene plasmons. This explains the relatively large reflection of graphene plasmons at nanometer-size gaps. Our results thus suggest that introducing nanoscale discontinuities, either by graphene patterning or by tailoring the substrate topography, opens up promising pathways for the development of ultracompact graphene-based plasmonic devices and circuits.

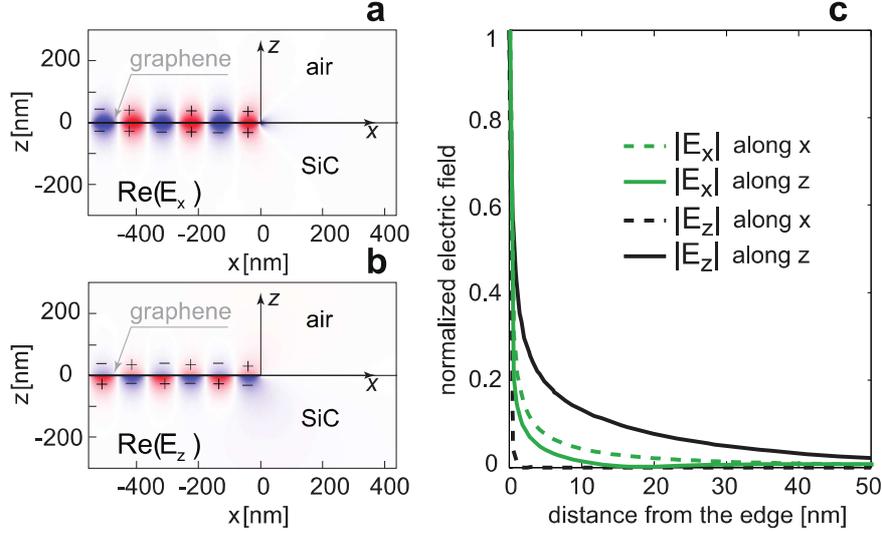

**Figure 4: Spatial distribution of the electric field of a plasmon plane wave reflecting at a graphene edge**. a) Instant snapshot of the x-component of the real part of the electric field $E_x$. b) The same as in (a), but for $E_z$. c), The dashed and solid curves indicate the decay of the fields along the x and z axes respectively. The fields are normalized to their value in the origin. The calculation is made for $|E_F| = 0.34$ eV and the light wavelength $\lambda_0 = 9.3$ μm, which corresponds to $\lambda_{SP} = 188$ nm.

Apart from its importance in plasmonics, this work has also implications in the field of epitaxial graphene. QFMLG has been shown to have advantages compared to regular epitaxial graphene on a buffer layer[14,16]. For example, the charge carrier mobility of QFMLG is almost independent of the temperature.[20] Buffer-layer-elimination by interface hydrogenation of epitaxial graphene was demonstrated to improve the device performance in graphene field effect transistors (FETs).[29] Here we have found direct evidence for the existence of discontinuities at step edges exceeding a critical height of about 1.5 nm. The absence of graphene overgrowth at large steps is in line with the observation that no buffer layer (and thus graphene after the hydrogen intercalation process) is formed during the graphitization of nonpolar surfaces[25] to which the step surface belongs. The graphene overgrowth of steps smaller than 1 nm might be explained by the formation of graphene bridges connecting the terrace surfaces, owing to diffusion of carbon atoms accross the steps. Why the formation of these graphene bridges starts at step heights around 1.5 nm and whether it is a specific property of QFMLG on SiC(0001) are open questions, which have to await future studies. We expect that the discontinuities in graphene have a strong impact on the electrical transport across the step edges[30] compared to the charge transport within the terraces.

**Acknowledgements**

RH acknowledges support by the ERC Starting Grant 258461 (TERATOMO) and the National Project MAT2012-36580 from the Spanish Ministerio de Ciencia e Innovación. M.L.N., A.Y.N., T.M.S. and L.M.M. acknowledge the Spanish Ministry of Science and Innovation grant MAT2011-28581-C02. ST and F.J.G.A. acknowledge support from the Spanish MEC (contract No. MAT2010-14885). The work of I.C. and A.B.K. was supported by Swiss National Science Foundation (grant No. 200020-140710). F.H.L.K. acknowledges support by the Fundacicio Cellex Barcelona, the ERC Career integration grant 294056 (GRANOP) and the ERC starting grant 307806 (CarbonLight). We acknowledge support by the EC under Graphene Flagship (contract no. CNECT-ICT-604391).

**Declaration of competing financial interests:**

R.H. is co-founder of Neaspec GmbH, a company producing scattering-type scanning near-field optical microscope systems. All other authors declare no competing financial interests.

**Methods**

Quasi-free-standing graphene was obtained on the silicon face of SiC by graphitization at 1450 $^{\circ}$C in Ar atmosphere. The first carbon layer (also called 'buffer layer'), initially covalently bonded to the substrate, was transformed into p-doped quasi-free standing graphene by hydrogen intercalation at 600 $^{\circ}$C[15,19,20]. This technique results in the average hole concentration close to $\sim 6\times 10^{12}$ cm$^{-2}$, corresponding to a Fermi energy $E_F \approx -0.3$ eV with respect to the Dirac point.[22] The charge mobility at room temperature in similar samples was found to be close to 1500 cm$^2$/(V·s)[19].

# Supporting Information

J. Chen, M. L. Nesterov, A. Yu. Nikitin, S. Thongrattanasiri, P. Alonso-González, T. M. Slipchenko, F. Speck, M. Ostler, Th. Seyller, I. Crassee, F. Koppens, L. Martin-Moreno, F. J. García de Abajo, A.B. Kuzmenko, R. Hillenbrand

In this document, we provide a detailed description of the numerical technique used to model the experimental data presented in the main text. We perform finite-element calculations (using COMSOL software), of the near-field profiles and directly compare them with the experimental profiles extracted from the s-SNOM measurements.

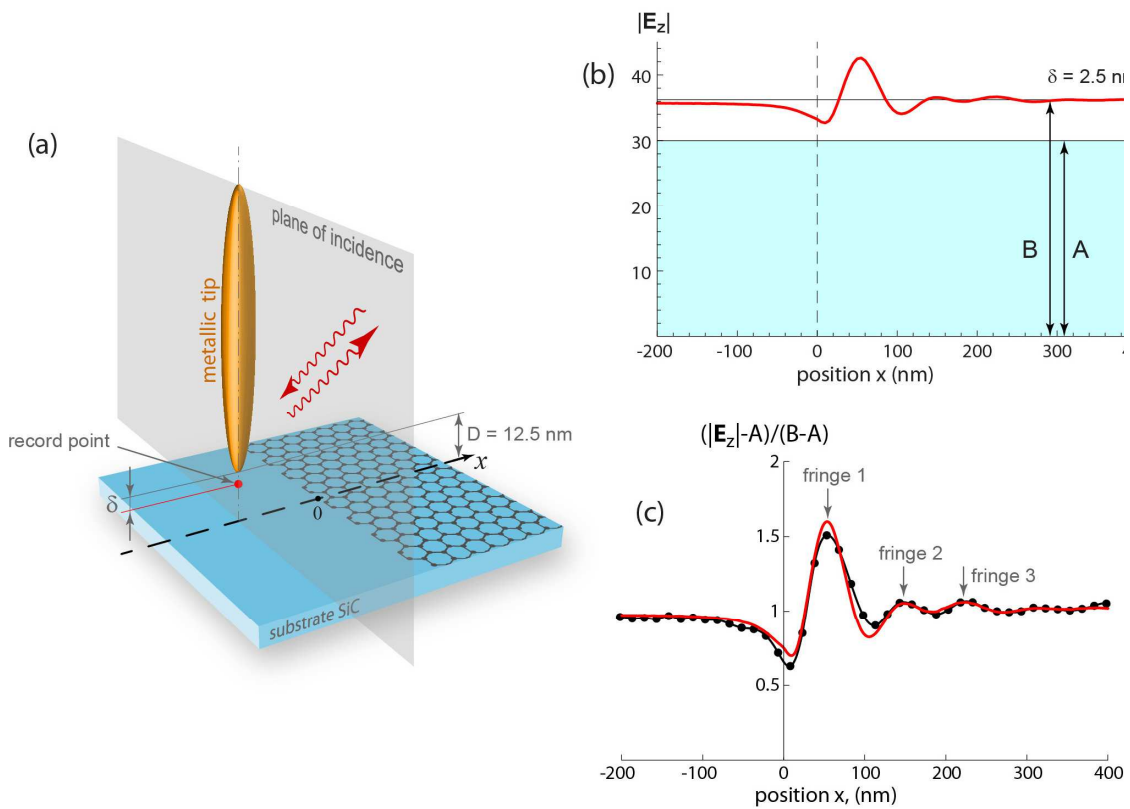

Fig. S1. (a) Schematic of the model used in the simulations. (b) The near-field profile corresponding to the graphene edge. (c) Comparison between the model and experiment. The red curve presents our theoretical result, whereas the black curve with dots shows the experimentally measured profile. The parameters taken for the simulations are as follows: wavelength 9.3 μm, SiC refractive index $n_{SiC}$ = 1.684 + 0.0116i, Fermi level $E_F$ = -0.34 eV, and relaxation time $\tau$ = 0.05 ps.

A calculation which fully takes into account the sophisticated experimental setup (including an oscillating tip and the far-field demodulated signal[1]) is not feasible with currently available computers. For this reason, instead of calculating the oscillation of the tip and the demodulation of the signal, we consider a simplified system described by a few parameters as explained below.

The model geometry consists of a 3D ellipsoidal metallic tip placed in the vicinity of the structure under study (see schematic in Fig. S1a). We illuminate the structure with a p-polarized plane wave at an angle of incidence of 45 degrees, so that its electric field has a finite component parallel on the tip. The tip is raster scanned along the x-direction at the fixed vertical distance D from the structure surface (being from the physical point of view an "average" distance for the real tip in the experiment). We record the z-component of the near field (NF), $|E_z|$, at the point located right below the tip apex, at a distance δ. Since the experimentally observed signal presents the field scattered by the tip, we take the recording point close to the tip termination. The black dashed curve in Fig. S1 (a) displays the scan path of the tip across a graphene edge, which is used as a reference system. We record a set of NF profiles (NFPs) $|E_z(x)|$ for different values of D and δ. After that, we select a profile which provides the best fitting to the experimental background-free signal $s_4(x)$. The fitting consists in the following linear transformation (see Fig.S1 (b)):

$$\frac{s_4(x)}{s_G} = \frac{|E_z(x)| - A}{B - A}, \qquad (A1)$$

where $s_G$ is the measured signal and B is the calculated field amplitude, both in the graphene region of the sample far away from the defects. The parameter A takes into account an average background signal which disappears in the experiment due to the demodulation. We have to subtract this average signal because in the simulations we do not take the demodulation explicitly into account. Notice that the tip-sample separation D affects the value of the constant B in Eq. (A1) and also the visibility of the interference fringes (see Fig. S2a). Figure S2b illustrates that the shape of the NFPs does not change much with the displacement of the recording point $\delta$, while the level B does.

A comparison between theoretical and experimental NFPs is shown in Fig. S1c. One can see that the shape of the NF is perfectly captured by the model. A point to note is that the optical conductivity of graphene in these calculations is assumed to be independent of coordinate $x$, i.e. $\sigma(x) = const$.

We find that the best combination for the studied sample is a graphene-tip separation of 12.5 nm and a NFP recording point position placed 2.5 nm below the ellipsoid, so that D = 12.5 nm and δ = 2.5 nm in Fig. S1. Once the constants A, D, and δ are found from this fitting, they are kept unchanged in the analysis of the other studied defect structures.

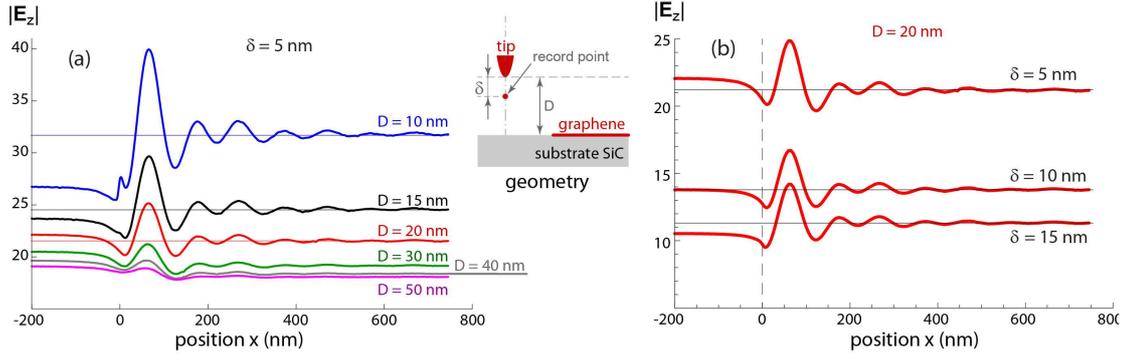

Fig. S2. Near-field profiles for different tip-sample separations D and positions δ of the point below the tip where the field are recorded. In (a) δ = 5 nm, while in (b) D= 20 nm. The parameters taken for the simulations are as follows: wavelength 9.2 μm, SiC refractive index $n_{SiC}$ = 1.7 + 0.0127i, Fermi level $E_F$=-0.4 eV, and relaxation time τ = 0.1 ps.

It is interesting to mention that the generated NFPs are not very sensitive to the shape of the elongated object (i.e., similar results are obtained for a cone, an elongated ellipsoid, etc.), which affects mainly the saturation constant B. As an example, we demonstrate in Fig. S3 that the NFPs for the cone- and spheroid-shaped tips coincide after performing the procedure specified in the r.h.s of Eq. (A1). Important requirements for the simulated tip should be its size (much larger than the studied defects) and the aspect ratio between longitudinal and transversal lengths. These two constraints guarantee the correct enhancement factor and sharpness of the tip termination. For all the simulations presented both in the manuscript and in this Supporting Information, we use an elongated ellipsoid of 900 nm in length and 100 nm in diameter.

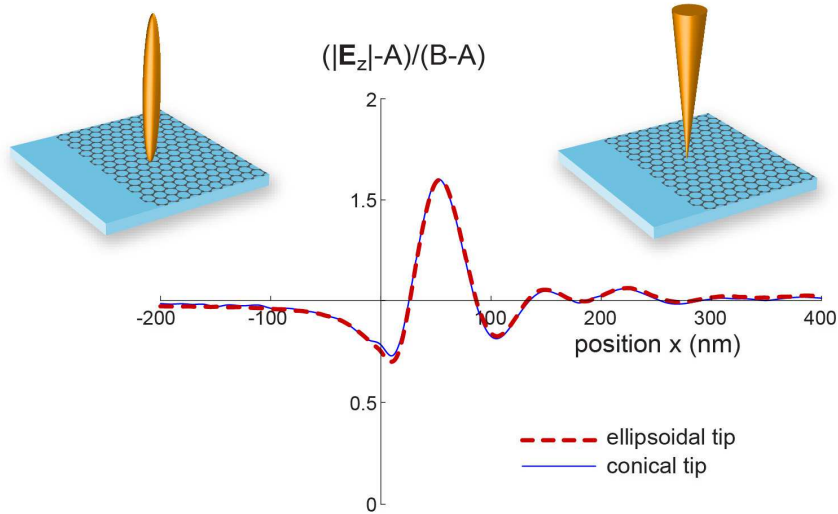

Fig. S3. Near field profiles calculated using the conical and spheroidal shapes of the tip for the graphene edge. The parameters taken for the simulations are the same as in Fig. S1.